\documentclass[10pt,reprint]{iopart}

\usepackage{graphicx}
\usepackage{psfrag}
\usepackage{color}
\usepackage{amsbsy}
\usepackage{amssymb}
\usepackage{iopams}

\def\dd{\mbox{d}}
\def\a{\alpha}
\def\b{\beta}
\def\g{\gamma}
\def\d{\delta}
\def\ve{\varepsilon}

\def\s{\sigma}

\def\s{\sigma}

\begin{document}

\title{Hydrodynamic mean field solutions of 1D exclusion
processes with spatially varying hopping rates}

\title[]{}

\author{Greg Lakatos, John O'Brien, and Tom Chou} 

\address{Dept. of Biomathematics and Institute for Pure and Applied
Mathematics, UCLA, Los Angeles, CA, 90095, USA}

\begin{abstract}
We analyze the open boundary partially asymmetric exclusion process
with smoothly varying internal hopping rates in the infinite-size,
mean field limit. The mean field equations for particle densities are
written in terms of Ricatti equations with the steady-state current
$J$ as a parameter. These equations are solved both analytically and
numerically. Upon imposing the boundary conditions set by the
injection and extraction rates, the currents $J$ are found
self-consistently. We find a number of cases where analytic solutions
can be found exactly or approximated. Results for $J$ from asymptotic
analyses for slowly varying hopping rates agree extremely well with
those from extensive Monte Carlo simulations, suggesting that mean
field currents asymptotically approach the exact currents in the
hydrodynamic limit, as the hopping rates vary slowly over the lattice.
If the forward hopping rate is greater than or less than the backward
hopping rate throughout the entire chain, the three standard
steady-state phases are preserved.  Our analysis reveals the
sensitivity of the current to the relative phase between the forward
and backward hopping rate functions.
\end{abstract}

\ead{tomchou@ucla.edu}



\section{Introduction} 

\begin{figure}
    \begin{center} 
         \includegraphics[height=1.5in]{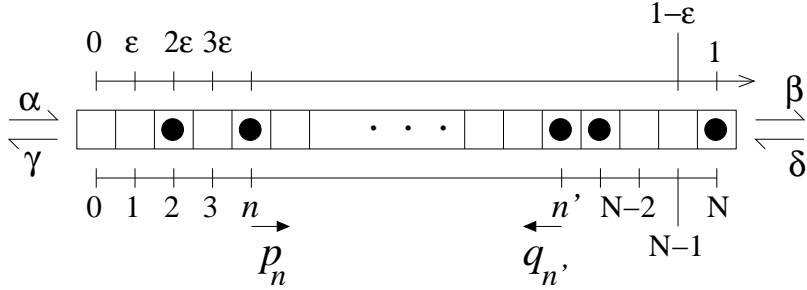}
      \end{center} 
\caption{The $N+1$-site open boundary,
      partially asymmetric exclusion process.  The continuum limit is
      a taken by setting each lattice site to 
size $\ve = 1/N$, thereby normalizing normalizing the total length.} 
\label{LATTICE}
\end{figure}

Asymmetric exclusion processes (ASEP) have been used as model
nonequilibrium statistical mechanical systems to represent many
physical processes such as traffic flow
\cite{TRAFFIC,TRAFFIC2,TRAFFIC3}, ion transport across channels
\cite{CHOU98,CHOU99}, mRNA translation \cite{MRNA1,MRNA2,MRNA3}, and
vesicle translocation along microtubules \cite{FREY}. For uniform
hopping rates, the steady-state currents, particle densities, and
correlations of the one-dimensional totally (TASEP) and partially
asymmetric exclusion process (PASEP) have been studied extensively
using recursion methods, exact matrix product techniques, and mean
field approximations \cite{DOMANY,DER93,DER98,SANDOW,RITT}. The PASEP
is described in Figure \ref{LATTICE}, and is comprised of
one-dimensional lattice of sites, each of which can only be empty of
singly occupied. The rules governing the dynamics of this system are
as follows: a particle at site $n$ hops to site $n+1$ with probability
$p_{n}\dd t$ in the time infinitesimal $\dd t$, only if site $n+1$ is
empty.  Similarly, it can hop backward to site $n-1$ (if site $n-1$ is
empty) with probability $q_{n}\dd t$. At the left and right boundaries
(sites $n=1$ and $n=N$, respectively) the injection probabilities are
$\a\dd t$ and $\delta \dd t$, respectively, provided these sites are
unoccupied. Extraction of particles from sites $n=1$ and $n=N$ occur
at rates $\gamma$ and $\beta$, respectively. Particles do not hop if
others are blocking their target sites. We will only consider the averaged
steady-state configurations of this system.

Exact steady-state currents for the case of constant $p_{n}=p$ and
$q_{n}=q$ have been found \cite{SANDOW,RITT}. In the case where $p\neq
q$, the exact solution in the infinite lattice limit ($N\rightarrow
\infty$) exhibits three phases described by maximal current, high
particle density, and low particle density. Within each of these
phases, the steady-state particle current is described by explicit
analytical expressions \cite{RITT}. Additional subphases corresponding
to different density {\it profiles} arise within the high and low
density current regimes \cite{KOLO2,NAGY}. When the forward and
backward hopping rates (out of and into each site) are equal, the
chain is purely diffusive and is driven only by a difference between
the injection/extraction rates at the boundaries.  In this case, only
a single, smooth (with respect to the injection/extraction rates)
current phase exists.

In many systems modelled by the PASEP, the internal hopping rates are
spatially varying.  For example, variations in the hopping rates may
arise in pores that have internal molecular structure, microtubules
tracks (on which molecular motors move) that are comprised of periodic
subunits, or from variations in mRNA or DNA sequence. Variations in
the forward hopping rate for fixed lattice defects in a TASEP have
been treated approximately in the limit of few, isolated defects
\cite{MRNA3,KOLO}, and in the periodic case where the forward hopping
rate takes on two values \cite{PERIODIC}.

In this paper, we consider spatially varying internal forward {\it
and} backward particle hopping rates in a PASEP.  We find solutions
for the current and density when the forward and backward hopping
rates are given by functions $p_{n}$ and $q_{n}$ that vary slowly with
the lattice position $n$. In the thermodynamic, mean field limit, the
equation of motion for the mean occupancy at each site can be
described in terms of a nonlinear continuum equation involving the
coarse grained mean occupations $\sigma(x)$, and the continuum hopping
rate functions $p(x)$ and $q(x)$. In the next section we derive the
{\it steady-state} continuum equations by expanding the occupancy
evolution equations in powers of $\ve = 1/N$, where $N\rightarrow
\infty$ is the total number of lattice sites in the chain.  We the
consider four general classes of the hopping functions $p(x)$ and
$q(x)$.  In Section 3, we treat the ``pure diffusion'' limit where
$p_{n} = q_{n+1}$, or, in the continuum limit, $p(x)=q(x+\ve)$. In
this limit, $q(x) \simeq p(x) -\ve p'(x)$, the mean-field equations
become linear, and exact simple results are recovered. In Section 4,
we consider the ``shifted diffusion'' limit, where the forward and
backward hopping rates at each site are identical in the sense that
$p_{n} =q_{n}$, or, $\vert p(x)-q(x)\vert \ll o(\ve)$. We find exact
implicit solutions for special forms of $p(x)$,. These results are
markedly different from those found for pure diffusion, although the
structure of $p(x)$ and $q(x)$ are nearly identical for the two cases.
The case where $\vert p(x) - q(x)\vert > O(\ve)$, and $p(x)-q(x)$ does
not change sign is considered in Section 5. Asymptotic analysis
indicates that the standard three phase structure found for constant
$p, q$ \cite{SANDOW,RITT} is preserved qualitatively.  In Section 6,
we show that internal density boundary layers arise if $p(x)-q(x)$
crosses zero for $0<x<1$.  This case eluded analytic treatment so only
numerical and simulation results were obtained.  In all cases, we
compare our results with numerics and continuous time Monte-Carlo
simulations. In the Summary and Conclusions, we discuss the limits in
which one would expect mean-field approaches to yield exact
steady-state currents.

\section{Continuum Mean Field Limits}


Consider a one-dimensional lattice (Fig. 1) containing $N+1$ sites
each of length $\varepsilon$.  For the interior sites, the continuum
limit of this lattice will be defined by a sampling of all relevant
quantities ({\it e.g.}, density) at the centers of each lattice site.
Density profiles in the presence of sources and sinks exhibit rich
shock behavior as studied by Parmeggiani {\it et al.} \cite{FREY2},
and by Evans {\it et al.} \cite{EVANS}. Here, we will neglect
adsorption/desorption at the interior sites; however, we allow the
internal hopping rates to vary slowly along the chain.

The equation for the discrete occupation variable 
$\hat{\s}_{n} \in (0,1)$ in the chain interior is

\begin{equation}
\begin{array}{ll}
\displaystyle {\dd \hat{\s}_{n} \over \dd t} & =
(\hat{J}_{n-1}^{+}-\hat{J}_{n}^{+})+(\hat{J}_{n+1}^{-}-\hat{J}_{n}^{-}),
\quad 1\leq n \leq N-1,
\label{HATFIRST}
\end{array}
\end{equation}

\noindent where 

\begin{equation}
\hat{J}_{n}^{+} = p_{n}\hat{\s}_{n}(1-\hat{\s}_{n+1}) \quad \mbox{and}\quad 
\hat{J}_{n}^{-} =
q_{n}\hat{\s}_{n}(1-\hat{\s}_{n-1})
\label{JEXPAND}
\end{equation}

\noindent are the currents from site $n$ to site
$n+1$ and from site $n$ to site $n-1$, respectively. 

The mean field assumption implies that the ensemble averaged occupancies are 
uncorrelated, $\langle \hat{\s}_{n}\hat{\s}_{m}\rangle \approx 
\s_{n}\s_{m}$, where $\s_{n} \equiv \langle \hat{\s}_{n} \rangle$.
Upon taking an ensemble average of Eq. \ref{HATFIRST}, and 
applying the mean field approximation, the evolution equation for the 
mean occupancy in the chain interior $(1\leq n \leq N-1)$ becomes

\begin{equation}
\begin{array}{ll}
\displaystyle {\dd \s_{n} \over \dd t} & =
(J_{n-1}^{+}-J_{n}^{+})+(J_{n+1}^{-}-J_{n}^{-}) \\[13pt] \: &
\displaystyle \approx \ve {\partial \over \partial x}(J^{-}-J^{+}) +
{\ve^{2}\over 2}{\partial^{2} \over \partial x^{2}} (J^{+}+J^{-}).
\label{FIRST}
\end{array}
\end{equation}

\noindent where $J_{n}^{+} \equiv \langle \hat{J}_{n}^{+}\rangle 
= p_{n}\s_{n}(1-\s_{n+1})$ and 
$J_{n}^{-} \equiv \langle \hat{J}_{n}^{-}\rangle = q_{n}\s_{n}(1-\s_{n-1})$.
Upon extrapolating the continuous function according to
$\s(x=n\ve) = \s_{n}$, and Taylor expanding Eq. \ref{FIRST} in powers
of $\ve$, we find the continuum mean field equation:

\begin{equation}
\fl \hspace{1.5cm}  {\dd \s(x) \over \dd t} = \ve \left[(q-p)\s(1-\s)\right]' +
{\ve^{2}\over 2}\left[\left[(p+q)\s\right]'(1-\s) +
(p+q)\s\s'\right]'.
\label{INTERIOR}
\end{equation}

\noindent Assuming the steady-state limit, and integrating the conservation 
Eq. \ref{INTERIOR}, we obtain
\cite{FOOT1}

\begin{equation}
(p-q)\s(1-\s)  
- {\ve \over 2} \left(\left[(p+q)\s\right]'(1-\s)+ (p+q)\s\s'\right) = J,
\label{INTEGRATED1}
\end{equation}

\noindent where the integration constant $J$ is the steady-state
current.  Equation \ref{INTEGRATED1} can be rewritten in the Riccati
form

\begin{equation}
\ve \s'(x) = -JP(x) + Q(x)\s(x)(1-\s(x)),
\label{INTEGRATED2}
\end{equation}

\noindent where 

\begin{equation}
\begin{array}{l}
\displaystyle P(x) = {2 \over (p+q)} \quad \mbox{and} \\[13pt] 
\displaystyle Q(x) = \left[{2(p-q) \over (p+q)} 
- \ve {(p+q)'\over (p+q)}\right].
\end{array}
\end{equation}

\noindent The boundary densities are found by measuring the steady-state current
into and out of the first and last sites: $J = \a(1-\s_{0})-\g\s_{0}$
and $J = \b\s_{N}-\d(1-\s_{N})$, from which we find

\begin{equation}
\s_{0} = {\a- J \over \a + \g}\quad \mbox{and} \quad \s_{N} = {\d + J
\over \b + \d}.
\label{S0J}
\end{equation}








Equations \ref{INTEGRATED2} and \ref{S0J} form the basis
of our steady-state analysis.  Integrating Eq. \ref{INTEGRATED2} from
$x=0$ to $x=1$, and imposing the boundary conditions
(Eqs. \ref{S0J}), implicitly determines $J$. Once the
steady-state current $J$ is fixed, the mean field density profiles
are determined.  For certain $p(x), q(x)$, one may be able to
solve Eq. \ref{INTEGRATED2} analytically, and use this result along
with the boundary conditions \ref{S0J} to find $J$ in
closed form.

\section{Pure diffusion: $p(x)=q(x+\ve), \,0<x<1$}

Consider the special case $p_{n} = q_{n+1},\,\, 1\leq n \leq N-1$
where the hopping rates between  two sites are equal.  In this case,
there is no driving force on the particles and a net current arises
only from differences in injection and extraction rates at the two
ends. The quadratic terms in Eq. \ref{HATFIRST} cancel, the equation
for $\s_{n}$ becomes linear, and the mean field approximation is
exact. In the continuum approximation  $p(x)=q(x+\ve)$, 

\begin{equation}
\fl P(x) = {1\over p(x)}\left[1+{\ve \over 2}{p'\over p} +
{\ve^{2}\over 4}\left({p'^{2} \over p^{2}}-{p''\over p}\right)^{2}+O(\ve^{3})\right] \,\,\mbox{and}\,\, Q(x) = O(\ve^{3}),
\end{equation}


\noindent and to lowest order in $\ve$,

\begin{equation}
\s'(x) \simeq -{J \over \ve p(x)}.
\label{SIMPLE}
\end{equation}

\noindent Integration of Eq. \ref{SIMPLE} yields 

\begin{equation}
\s(1) - \s(0) \approx  -{J\over \ve}\int_{0}^{1} {1\over p(x')}\dd x' 
\equiv -{J\over \ve}\langle 1/p\rangle.
\end{equation}

\noindent Upon applying Eqs. \ref{S0J}, 
(with $q_{1}=p_{0}$ and $q_{N} = p_{N-1}$), and solving for $J$,

\begin{equation}
J = {(\a\b-\g\d) \over \langle 1/p\rangle (\a+\g)(\b+\d)
+(\a+\b+\g+\d)}.
\label{JDCONT}
\end{equation}

\noindent This same ``homogenization'' result, with the Riemann equivalent  

\begin{equation}
\langle 1/p\rangle \equiv \sum_{n=0}^{N-1}p_{n}^{-1},
\end{equation}

\noindent is also easily obtained by recursively solving the exact
discrete equation $p_{n-1}(\s_{n}-\s_{n-1}) = -J, \, (1\leq n \leq
N)$.  The corresponding density profile is obtained through

\begin{equation}
\s_{n} = \s_{0} + J\sum_{j=0}^{n-1}p_{j}^{-1}.
\label{SIGMAEXACT}
\end{equation}

\noindent For constant $p_{n}=q_{n}=p$, Eq. \ref{JDCONT} reduces to
trivial result for the steady-state current of an $N+1$ site
boundary-driven chain \cite{CHOU98,RITT}

\begin{equation}
J = {p(\a\b-\g\d) \over N(\a+\g)(\b+\d) + p(\a+\b+\g+\d)}.
\end{equation}










\section{Shifted diffusion: $p(x) \simeq q(x), \, 0<x<1$}

If $p_{n}\simeq q_{n}$, such that $\vert p(x) - q(x)\vert \ll o(\ve)$,
particles at each site hop equally to the right or the left, with
possibly different rates from site to site.  This case corresponds to
particles that cannot distinguish forward from backward motion but, as
detailed balance is violated, is not equivalent to pure diffusion.  We
will see that this slight change in the hopping rate structure from
the $p_{n} = q_{n+1}$ case results in a very different steady-state
current.

To lowest order, when  $p(x) \simeq q(x)$,

\begin{equation}
\displaystyle P(x) \simeq {1 \over p(x)} \quad \mbox{and}\quad  
Q(x) \simeq -\ve {p'(x) \over p(x)}.
\end{equation}

\noindent Since $Q(x)$ is of order $P(x)$, the $\s(1-\s)$ term 
in Eq. \ref{INTEGRATED2} cannot be neglected and, unlike the 
purely diffusive case, the problem is nonlinear. Therefore, 
we would not expect the mean field particle densities or currents to be 
necessarily exact.

In this case, there are various variable transforms that render the
Riccati equation analytically tractible.  The simplest case is where
the Riccati equation is separable. This occurs when $(p+q)'=$constant,
which implies $p(x)=q(x) = ax+b$.  Equation \ref{INTEGRATED2} can then
be integrated from $x=0$ and $\s = \s(0)= \s_{0}$ to give

\begin{equation}
\int_{0}^{x} {\dd x' \over ax'+b} = 
{1\over a}\int_{\s_{0}}^{\s(x)}{\dd \s \over (\s-\s_{+})(\s-\s_{-})},
\label{INT2}
\end{equation}

\noindent where $\s_{\pm} = 1/2 \pm 1/2\sqrt{1+4J/(\ve a)}$. 
Integrating (\ref{INT2}), we find the density profile
$\s(x)$  from

\begin{equation}
\left({ax+b \over b}\right)^{\s_{+}-\s_{-}} = 
\left({\s(x)-\s_{+}\over \s(x)-\s_{-}}\right)
\left({\s_{0}-\s_{-}\over \s_{0}-\s_{+}}\right).
\label{DENSITY1}
\end{equation}

\noindent An implicit formula for $J$ (to order $\ve$) is found by imposing
the boundary condition at $x=1$
($\s(1)=\s_{N} \approx \d/(\b+\d)$):

\begin{equation}
\left({a+b\over b}\right)^{\sqrt{1+4J/(\ve a)}} = 
\left({\d-(\b+\d)\s_{+}\over \d-(\b+\d)\s_{-}}\right)
\left({\a-(\a+\g)\s_{-}\over \a-(\a+\g)\s_{+}}\right).
\label{IMPLICIT1}
\end{equation}

\begin{figure}[htb]
    \begin{center} 
         \includegraphics[height=3.7in]{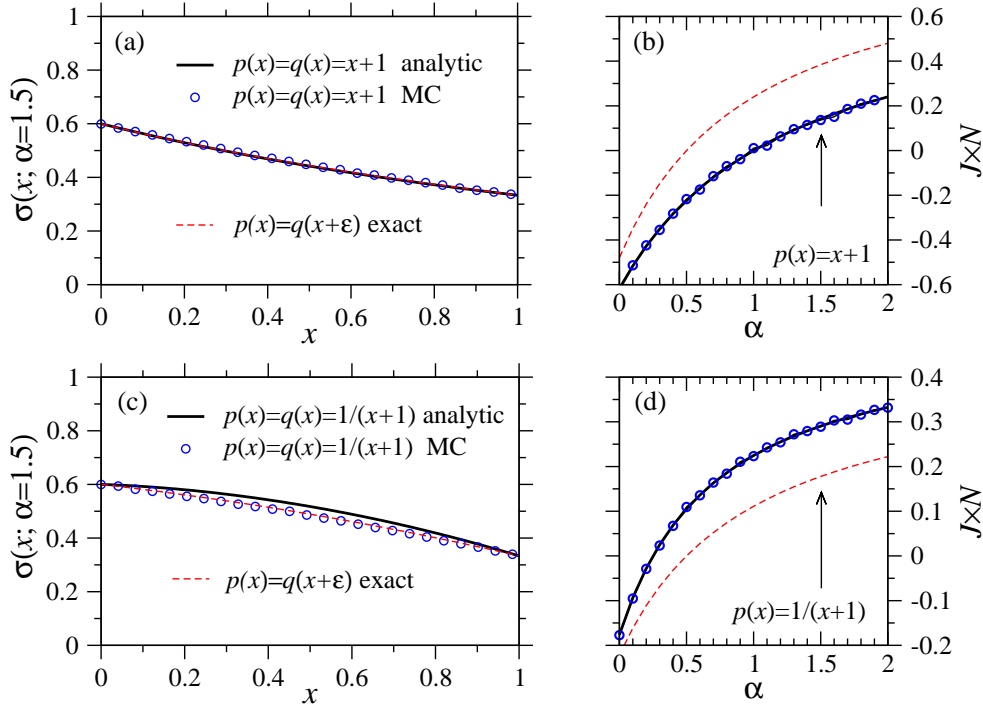}
      \end{center} 
\caption{Densities and currents from analytic solutions of the Riccati
equation and from simulations.  (a) The density profile resulting from
the linear hopping rate function $p(x)= q(x) = x+1$.  Results from the
analytic solution Eq. \ref{DENSITY1} (solid curve) and from
Monte-Carlo simulations (circles) are shown. Also shown is the exact
density for the purely diffusive case $p(x) = q(x+\ve) = x+1$ (dashed
curve).  (b) Currents derived from both the analytic solution
Eq. \ref{IMPLICIT1} (solid curve) and Monte-Carlo simulations
(circles). Also shown for contrast is the exact current in the purely
diffusive case (dashed curve).  (c) The density profiles associated
with the hopping rate function $p(x)=q(x)=1/(x+1)$. The solid,
circled, and dashed curves correspond to analytic, Monte-Carlo, and
exact diffusion (for $p(x) = q(x+\ve) = 1/(x+1)$) solutions.  (d)
Currents derived from both simulation and the analytic solution to
Eq. \ref{IMPLICIT2}. The arrows in (b) and (d) mark the value $\a=1.5$
used in plotting the density profiles shown in (a) and (c).}
\label{FIG2}
\end{figure}

\noindent The solution to Eq. \ref{IMPLICIT1} is found numerically and plotted in 
Fig. \ref{FIG2} for representative parameters.

Next, consider another analytic solution found by using the definition

\begin{equation}
\s(x) = {\ve\over Q(x)}{y'(x)\over y(x)}
\label{SY}
\end{equation}

\noindent which transforms Eq. \ref{INTEGRATED2} to

\begin{equation}
\fl Q(x)y''(x) - \left[Q'(x)+\ve^{-1}Q^{2}(x)\right]y'(x)+ \ve^{-2}JP(x)Q^{2}(x)y(x) = 0.
\label{LINEAR1}
\end{equation}

\noindent Provided 

\begin{equation}
Q'(x)+\ve^{-1}Q^{2}(x) = 0,
\label{Q2}
\end{equation}

\noindent and $Q(x) \neq 0$, we find

\begin{equation}
y''(x) +\ve^{-2}JP(x)Q(x)y = 0.
\label{Y2}
\end{equation}

\noindent The condition Eq. \ref{Q2} is solved by 

\begin{equation}
Q(x)= {\ve\over x+b} \equiv {2(p-q) \over (p+q)} - {\ve(p+q)'\over (p+q)},
\label{QPQ}
\end{equation}

\noindent which constrains $q(x)$ to $p(x)$ through 

\begin{equation}
q(x) = {e^{-2x/\ve} \over x+b}\left[\int^{x}(x'+b)g(x')e^{2x'/\ve} \dd
x' + \mbox{constant}\right],
\label{QP}
\end{equation}

\noindent where 

\begin{equation}
g(x) = \left({2\over \ve}-{1\over x+b}\right)p(x)-p'(x).
\end{equation}

\noindent Given pairs of $p(x), q(x)$ that satisfy Eq. \ref{QPQ} or
\ref{QP}, one can find analytic solutions to Eq. \ref{Y2},
reconstruct $\s(x)$ via Eq. \ref{SY}, and impose the boundary
conditions on $\s(0)$ and $\s(1)$ (Eqs. \ref{S0J}) to
find an implicit equation for $J$.  Note that the constraint
Eq. \ref{QP} allows for analytic solutions of Eq. \ref{INTEGRATED2}
for hopping rate functions more general than $p(x)=q(x)$. Restricting
ourselves to $p(x) = q(x)$, the only solution for $p(x)+q(x) = 2p(x)$
that satisfies Eq. \ref{QPQ} is

\begin{equation}
p(x) =a/(x+b).  
\end{equation}

\noindent Equation \ref{Y2} then becomes

\begin{equation}
y''(x) +k^{2} y(x)=0,\quad k^{2} = {J \over \ve a}
\end{equation}

\noindent admitting a solution of the form

\begin{equation}
y(x) \propto  e^{ikx} + ce^{-ikx}.
\label{Y}
\end{equation}

\noindent Upon setting $\s(0) = Q^{-1}(0)(y'(0)/y(0)) = \s_{0} = (\a-J)/(\a+\g)$,

\begin{equation}
c = {kb+ i\s_{0}\over kb-i\s_{0}}.
\label{C}
\end{equation}

\noindent Substituting  Eq. \ref{C} into Eqs. \ref{Y} and \ref{SY}, we find

\begin{equation}
\s(x) = (x+b)k{\s_{0}\cos kx - kb\sin kx \over \s_{0}\sin kx +kb \cos kx}.
\end{equation}

\noindent Finally, imposing the 
boundary condition at $x=1$ implicitly determines $J$:

\begin{equation}
\s(1) = k(b+1){\s_{0}\cos k - kb\sin k \over \s_{0}\sin k +kb\cos k} = \s_{N} = 
{\d+J \over \b+\d}.
\label{IMPLICIT2}
\end{equation}

\noindent Since $J \sim \ve a$, $\s_{0} \approx \a/(\a+\g)$ and
$\s_{N} \approx \d/(\b+\d)$ can be used to numerically solve
Eq. \ref{IMPLICIT2} for currents and densities.  Expanding $J\approx
0$ also shows that $J\propto \ve a (\a\b(b+1)-b\g\d)$.

In Figs. \ref{FIG2}, we plot (a) the densities and (b) the currents
for the hopping rate profile $p(x) = q(x) = ax+b$ as a function of
driving $\a$.  The results of extensive Monte-Carlo simulations using
the BKL continuous time algorithm \cite{BKL}, for a lattice of size
$N=1000$, are also shown. Both analytic (in the $\ve \rightarrow 0$
limit) and simulation results agree to a high degree of accuracy.  In
Figs. \ref{FIG2}(c) and (d), we plot the densities and currents
corresponding to the inverse hopping rate profile $p(x) = q(x)
=a/(x+b)$. Here, the current also agrees well with the Monte-Carlo
simulations.  However, there is a small discrepancy between the
densities from mean field theory and those from Monte-Carlo
simulations.  This discrepancy is not unexpected since correlations
are neglected in mean field theory. Also shown for comparison are the
densities and currents for the purely diffusive chain where $p(x) =
q(x+\ve)$.  These densities (dashed curves) are very close to those
corresponding to $p(x) = q(x)$, however, the diffusive currents are
significantly different. By shifting the backward hopping rate
function by $\ve$, the density profile changes only slightly. However,
since the steady-state currents scale as $\ve$, small changes in
boundary densities can lead to large relative differences in the
steady-state currents.

\section{Completely driven chain: $\vert p(x)-q(x)\vert \gg O(\ve), \, 0<x<1$}

In this Section, we consider significantly different forward and
backward hopping rates, and for simplicity, first assume $p(x) > q(x)$
for $0< x < 1$.  In this case, neither $P(x)$ nor $Q(x)$ is small, but
Eq. \ref{INTEGRATED2} can be treated using singular perturbation
theory and the appropriate implementation of density boundary layers.
Suppose a boundary layer arises near $x \sim \ve$.  Rescaling $x =\ve
y$, we find

\begin{equation}
{\dd \s(y) \over \dd y} = -J P(\ve y) + Q(\ve y)\s(y)(1-\s(y)).
\label{BL1}
\end{equation}

\noindent Within the boundary layer, $y \sim O(1)$, 
$P(\ve y) \approx P(0)$, and  $Q(\ve y) \approx Q(0)$. 
Equation \ref{BL1} can be integrated to find the left  inner solution

\begin{equation}
\fl \s^{in}_{\ell}(y) = {\s_{+}(0)(\s_{0}-\s_{-}(0))
-\s_{-}(0)(\s_{0}-\s_{+}(0))e^{(\s_{-}(0)
-\s_{+}(0))Q(0)y} \over 
\s_{0}-\s_{-}(0) - (\s_{0}-\s_{+}(0))e^{(\s_{-}(0)
-\s_{+}(0))Q(0)y}},
\end{equation}

\noindent where $\s_{\pm}(0)$ is also the outer solution 
to Eq. \ref{BL1},

\begin{equation}
\s_{\pm}(x) = {1\over 2} \pm {1 \over 2}\sqrt{1-{4JP(x) \over Q(x)}},
\end{equation}

\noindent evaluated as $x\rightarrow 0$. Of the two possible outer solutions,
only $\s_{+}(0)$ can match the inner solution
$\s^{in}_{\ell}(y\rightarrow \infty)$.  The uniform solution with a
density boundary layer at $x \sim \ve$ is thus

\begin{equation}
\s_{\ell}(x) = \s_{\ell}^{in}(x/\ve) + \s_{+}(x) - \s_{+}(x=0).
\label{SIGMAL}
\end{equation}

\noindent This solution automatically satisfies the boundary condition
at $x=0$: $\s_{\ell}(0) = \s_{0}$. The current is determined by
satisfying the boundary condition at $x=1$; and since 
$\s^{in}_{\ell}(y=1) \sim 
\s_{+}(0)$,

\begin{equation}
\begin{array}{rl}
\s_{\ell}(1) \simeq  & \displaystyle \s_{+}(1) = {1\over 2}+{1\over
2}\sqrt{1-{4JP(1)\over Q(1)}} \\[13pt] \: = & \displaystyle \s_{N}=
{\d + J \over \b+\d}.
\label{LAK1}
\end{array}
\end{equation}

\noindent The only possible solution to Eq. \ref{LAK1} is

\begin{equation}
\fl J = {1 \over 2}\left(\b-\d-{P(1)\over Q(1)}(\b+\d)^2\right) + 
{1\over 2}\sqrt{\left(\b-\d-{P(1)\over Q(1)}(\b+\d)^2\right)^{2}+4\b\d}.
\label{JH}
\end{equation}

In addition to this result, two other solutions to Eq. \ref{BL1}
exist. One with a boundary layer at $x\sim 1$, and another with
boundary layers at both $x\sim \ve$ and $x\sim 1$. If a boundary layer
exists only at $x\sim 1$, only $\s_{-}(x)$ can match the inner
solution near $x=1$ and the uniform the solution analogous to
Eq. \ref{SIGMAL} is

\begin{equation}
\s_{r}(x) = \s_{r}^{in}(x/\ve) + \s_{-}(x)-\s_{-}(x=1),
\end{equation}

\noindent where 

\begin{equation}
\fl \s_{r}^{in}(x/\ve) =  {\s_{+}(1)(\s_{N}-\s_{-}(1))
-\s_{-}(1)(\s_{N}-\s_{+}(1))e^{(\s_{+}(1)
-\s_{-}(1))Q(1)(1-x)/\ve} \over 
\s_{N}-\s_{-}(1) - (\s_{N}-\s_{+}(1))e^{(\s_{+}(1)
-\s_{-}(1))Q(1)(1-x)/\ve}}.
\end{equation}

\noindent In this case, the self-consistent current is found 
from $\s_{r}(0)= \s_{0}$:

\begin{equation}
\fl J = {1\over 2}\left(\a-\g-{P(0)\over Q(0)}(\a+\g)^2\right)+{1\over 2}
\sqrt{\left(\a-\g-{P(0)\over Q(0)}(\a+\g)^2\right)^{2}+4\a\g}.
\label{JL}
\end{equation}
 
When both boundary layers exist, the outer solutions must match at at
least one intermediate interior position $\s_{+}(x^{*}) =
\s_{-}(x^{*}), \, 0 \ll x^{*}\ll 1$. The corresponding
uniform solution is 

\begin{equation}
\s_{*}(x) = \bigg\{\begin{array}{c}\s_{\ell}(x) \quad 0 \leq x \leq
x^{*} \\[13pt] \s_{r}(x) \quad x^{*} \leq x \leq 1
\end{array}
\end{equation}

\noindent with corresponding maximal current

\begin{equation}
J_{max} = {Q(x^{*}) \over 4P(x^{*})}.
\label{JMAX}
\end{equation}

\begin{figure}[htb]
    \begin{center} 
         \includegraphics[height=2.8in]{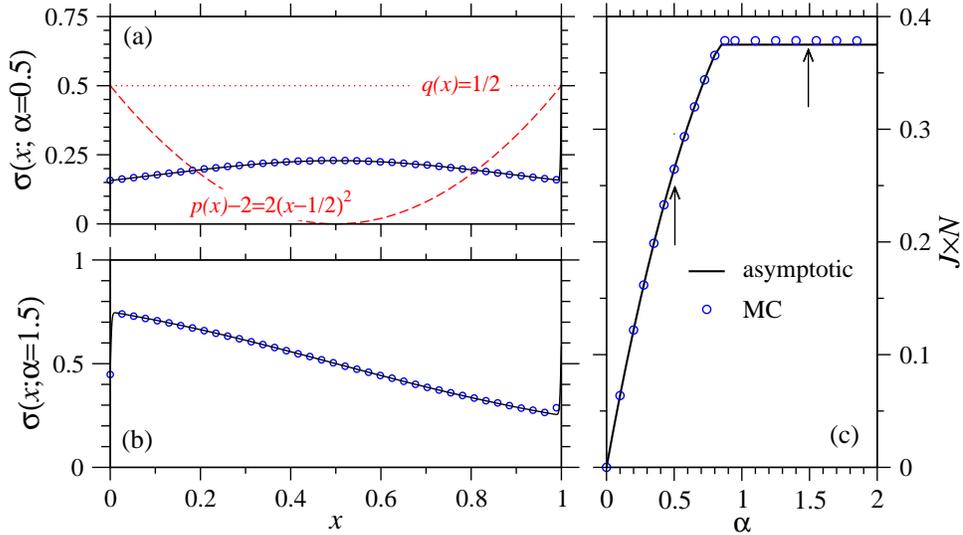}
      \end{center} 
\caption{Densities and currents from asymptotic solutions of the
Riccati equation and from simulations. Parameters used are
$\ve=1/1000$, $\beta=\gamma=1$, $\delta=0.5$, and $p(x)=2+2(x-1/2)^2 >
q(x) = 1/2$.  In all plots, the solid black curves correspond to
asymptotic solutions of the mean field equations, while the blue
circles correspond to results provided by Monte-Carlo simulations.
(a) Asymptotic and simulation densities for $\alpha=0.5$. These
parameters used render the system in a low density, entry rate-limited
regime. (b) For $\alpha=1.5$, the system is in the maximal current
phase ($J_{max}= 3/8$), where boundary layers arise at both $x\approx
0$ and $x\approx 1$, and $x^{*}=1/2$. (c). The steady-state current as
a function of $\alpha$. Given the other parameters used, the system
transitions from a low density to a maximal current phase at $\alpha =
5/6$. The parameters $\a=0.5, 1.5$ used in plotting the densities in
(a) and (b) are marked with arrows.}
\label{FIG3}
\end{figure}

The solutions Eqs. \ref{JH}, \ref{JL}, and \ref{JMAX} are valid only
in the parameter regimes where $J \leq Q(x^{*})/4P(x^{*})$ and $0 \leq
\s \leq 1$, and represent a generalization of the well-established
three phase current structure arising in the constant $p,q$ PASEP
\cite{SANDOW,RITT}.  This three phase structure is preserved only if
$p(x)- q(x)$ does not change sign on $x\in [0,1]$.  Note that if $p,
q$ are constant and $p-q>0$, the inner solutions are exact on $x\in
[0,1]$ and we recover the known results for the PASEP
\cite{SANDOW,RITT}. The mean field densities and currents for $p(x) =
2+2(x-1/2)^2$ and $q=1/2$ are plotted in Figure \ref{FIG3}, along with
results from continuous-time Monte-Carlo simulations. The agreement is
extremely good between the asymptotic mean field currents and
simulation currents, suggesting that the basic physics of the three
phase structure is preserved and that mean field theory provides
exact, steady-state currents. The densities are also in good
agreement, except in barely discernible region within the boundary
layers where mean field and simulation derived densities differ.  Note
that there is also a slight discrepancy between mean field and
simulation currents in the maximal current regime (Fig. 3c).  The
underestimation of the current by the mean field analysis results from
the finite size of the rate-limiting region. Although $\ve$ is small,
and $p(x)$ is reasonably slowly varying, the rate limiting region at
$x\approx 1/2$ is small enough for actual current (from MC
simulations) to be noticeably greater than the asymptotic mean field
result.

\section{Opposing drifts: $\vert p(x)-q(x)\vert \gg O(\ve)$ \mbox{except 
at countable points} $x_{0}$}

Finally, consider the important class of hopping rates where $\vert
p(x)-q(x)\vert \gg O(\ve)$, except at certain points $x_{0}$ where
$p(x)-q(x)$ crosses zero, like $Q(x)$ in the $\ve\rightarrow 0$ limit.
Examples of $p(x),q(x)$ with these properties are $p(x) = a+bx$, $q(x)
= a+b(1-x)$ (where $x_{0}=1/2$), and periodic $p(x), q(x)$ such that
$Q(x)$ oscillates above and below zero. Periodic hopping rates may
arise during transport through pores with atomic periodicity.  For
example, periodic arrangements of atoms or molecules within the pore
would impart a periodic potential on translocation of particles of the
form $p(x) \propto \exp[(V(x)-V(x+\ve))/k_{B}T]$ and $q(x) \propto
\exp[(V(x)-V(x-\ve))/k_{B}T]$, which are periodic if $V(x)$, the
interaction potential as a function of the coordinate $x$ along the
axis of the chain is itself periodic.

\begin{figure}[htb]
    \begin{center} 
        \includegraphics[height=3.7in]{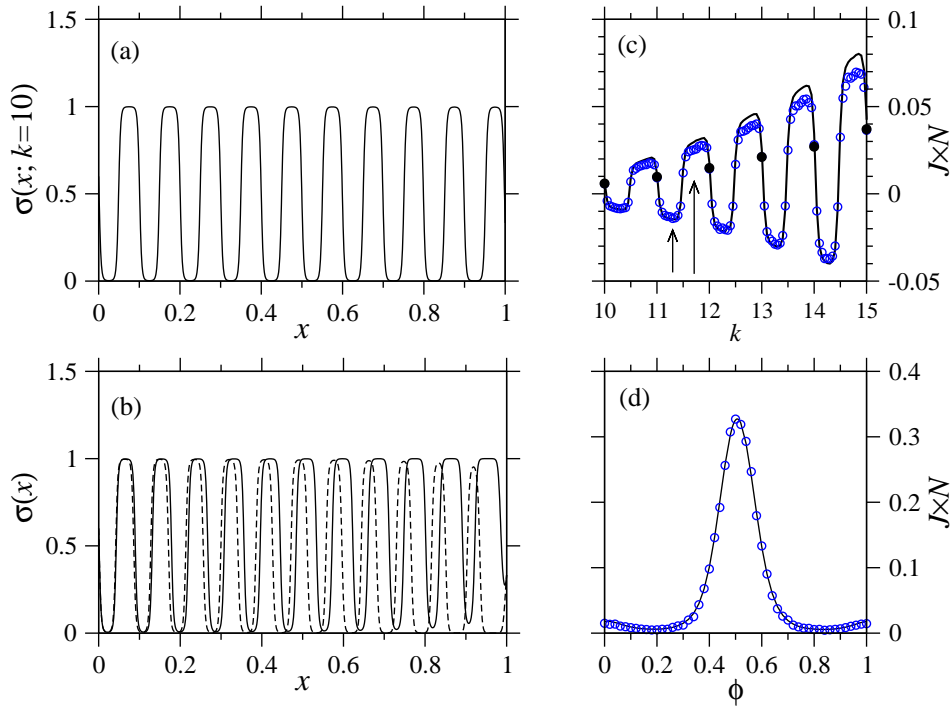}
      \end{center} 
\caption{Densities and currents from solutions of the Riccati equation
and from simulations for the periodic hopping rates modeled by
Eq. \ref{PERIODICPQ}. The parameters used are: $a=1$, $b=0.5$,
$\a=1.5$, $\b=\g=1$, $\d=0.5$. (a) $p(x)$ (dashed), $q(x)$
(dotted), and the numerically solved density profile for $k=10,
\phi=0$. (b) Numerically computed density profiles for $k=11.3,
\phi=0$ (solid) and $k=11.7, \phi=0$ (dashed). (c) Numerically solved
(solid) and simulation-derived (circles) steady-state currents as a
function of $k$ ($\phi=0$). The arrows indicate $k=11.3, 11.7$ used
for generating the profiles in (b). Currents associated with integer
values of $k$ are denoted by filled circles. (d) The current as a
function of the phase difference $\phi$ between $p(x)$ and $q(x)$ for
$k=12$.}
\label{FIG4}
\end{figure}

Instead of specifying a detailed, molecular model for the hopping
rates, we assume $p(x), q(x)$ to be functions that qualitatively
capture the physics arising from periodic hopping rates. The
qualitative dependence of the steady-state currents and densities
should not depend upon the exact, quantitative forms chosen for the
periodic hopping rates. Therefore, for simplicity, we assume:

\begin{equation}
\begin{array}{l}
\displaystyle p(x) = a+b\sin^{2}(\pi k x) \\[13pt] \displaystyle q(x) =
a+b\cos^{2}(\pi k x+\pi \phi),
\label{PERIODICPQ}
\end{array}
\end{equation}

\noindent with $k> 1$ and $a>b$. This functional form for the hopping
rates captures the periodicity of the pore potential and allows for a
phase difference $\phi$ between the forward and backward hopping
rates.  When $\phi=0$, $p(x)+q(x) = 2a+b$, $P(x) = 2/(2a+b)$, and the
function $Q(x) = -2b\cos (2\pi kx)/(2a+b)$ crosses zero at points
$x_{0} = (2n+1)/(4k)$.  Despite this simplification, there is no
analytic solution to Eq. \ref{INTEGRATED2}, and we were unable to find
approximations. Asymptotic analysis of the Riccati is also difficult
due to the existence of multiple, interior boundary layers, and the
fact that we must determine the boundary densities to $O(\ve)$ in
order to extract the current $J$. Moreover, numerical solutions to
Eq. \ref{INTEGRATED2} are difficult to obtain for extremely small
$\ve$ since numerical errors build up as one integrates
Eq. \ref{INTEGRATED2} from $x=0$ to $x=1$.  Nonetheless, we compare
currents and densities derived from numerics and continuous-time
Monte-Carlo simulations.

Figures 4 show numerically computed ($\ve = 0.001$) and simulated
($N=1001$) densities and currents for periodic hopping rates
Eq. \ref{PERIODICPQ}. In Fig 4(a) are the functions $p(x), q(x)$,
and the density profile for $k=10$ and $\phi=0$. Due to the
oscillatory nature of $p(x)$ and $q(x)$, the densities are locally
compressed and rarefied, rapidly jumping between $\s(x) = 0$ and
$\s(x)=1$. The numerically computed densities for noninteger $k$ are
also shown in Fig. 4(b). Small changes in $k$ can cause large
variations in the density near the $x = 1$ boundary, causing dramatic
changes in the current, as shown in Fig. 4(c). Fig 4(d) shows the
sensitivity of the current to variations in the phase $\phi$.  For
clarity have shown only numerically computed density profiles: in our
plots, densities found from simulations are nearly indiscernable from
those found numerically. As in all other cases of slowly varying
hopping rates, mean-field theory appears to yield exact steady-state
currents.  The agreement between numerical and simulated data is
extremely good in Fig. 4(c), but the discrepancy increases as the
number of hopping rate oscillations $k$ increases.

\section{Summary and Conclusions}

We have formulated the mean field approximation of a partially
asymmetric exclusion process in the hydrodynamic limit in terms of the
solution to the Riccati equation. This nonlinear equation can be
solved in special cases and asymptotically analyzed in others. We
compare numerical and analytical results with results from extensive,
($10^9 - 10^10$ steps) continuous time Monte-Carlo simulations and
find extremely good agreement for the steady-state particle currents.
The numerical simulations fall well within the within the simulation
error, typically $\lesssim 1\%$ and barely discernible in our plots.
This agreement holds for all hopping rate profiles considered,
provided they do not vary rapidly along the lattice.  Moreover,
although we have not proven that soutions to the Ricatti equations
(Eqs. \ref{INTEGRATED2}) yield exact currents, comparison of the
numerical solutions for the current with those obtained from extensive
continuous-time MC simulations shows a decreasing discrepancy as
$\ve/\ell \rightarrow 0$, provided sufficient long simulations are
performed.  Therefore, we conjecture that mean field approximations
provide {\it asymptotically} exact steady-state currents as long as
the hopping rate structure is smoothly varying in the thermodynamic
($N\rightarrow \infty$) limit.

The simulated densities, as expected, are quantitatively different
from those obtained from the numeric or analytic solution of the
Ricatti equation.  Moreover, we find that the three-phase current
structure of the PASEP is preserved when $p(x)>q(x)$. For cases where
$\vert p(x)-q(x)\vert \leq O(\ve)$ (pure diffusion and shifted
diffusion), $J\sim \ve$, but is sensitive to even slight shifts
between the functions $p(x)$ and $q(x)$. The cases $p(x) = q(x)$ and
$p(x) = q(x+\ve)$ correspond to physically realizeable systems, yet
yield very different results.  When $p(x)$ and $q(x)$ vary
periodically, as might be expected along a molecular channel
constructed from a periodic array of atoms or molecules, the currents
derived from solving Eq. \ref{INTEGRATED2} also appear to be exact,
provided there are a large number of sites in each period.

\vspace{5mm}

\noindent {\bf Acknowledgments} The authors are grateful for support
from the US National Science Foundation through grants DMS-0206733 and
DMS-0349195, and from the US National Institutes of Health through
grant K25AI058672. JO was also supported by NIGMS Systems and Integrative
Biology Training Grant 5T32GM008185.  This research has been enabled
by the use of WestGrid computing resources which are funded in part by
the Canada Foundation for Innovation, Alberta Innovation and Science,
BC Advanced Education, and the participating research
institutions. WestGrid equipment is provided by IBM, Hewlett Packard,
and SGI.



\end{document}